\documentclass[aps,showpacs,preprint,tightenlines,nofootinbib]{revtex4}
\usepackage[tbtags]{amsmath}
\usepackage{amsfonts,graphicx,bm}

\def\ts{\textstyle}
\def\beq{\begin{equation}}
\def\eeq{\end{equation}}
\def\eeql#1{\label{#1} \end{equation}}
\def\a{\alpha}
\def\D{\Delta}
\def\W{\Omega}
\def\c{\gamma}

\def\w{\omega}
\renewcommand{\O}{\mathcal{O}}

\def\hp{\mskip0.5\thinmuskip}
\def\im{\mathop{\rm Im}\nolimits}
\def\re{\mathop{\rm Re}\nolimits}
\newcommand{\To}{\rightarrow}

\begin{document}

\title{WKB analysis of the Regge--Wheeler equation down in the frequency plane}

\author{Alec \surname{Maassen van den Brink}}
\email{alec@dwavesys.com}
\affiliation{Physics Department, The Chinese University of Hong Kong, Hong Kong, China}

\date{\today}


\begin{abstract}
The Regge--Wheeler equation for black-hole gravitational waves is analyzed for large negative imaginary frequencies, leading to a calculation of the cut strength for waves outgoing to infinity. In the---limited---region of overlap, the results agree well with numerical findings [Class.\ Quantum Grav.\ \textbf{20}, L217 (2003)]. Requiring these waves to be outgoing into the horizon as well subsequently yields an analytic formula for the highly damped Schwarzschild quasinormal modes, \emph{including} the leading correction. Just as in the WKB quantization of, e.g., the harmonic oscillator, solutions in different regions of space have to be joined through a connection formula, valid near the boundary between them where WKB breaks down. For the oscillator, this boundary is given by the classical turning points; fascinatingly, the connection here involves an expansion around the black-hole singularity $r=0$.
\end{abstract}


\maketitle


\section{Introduction}

Black-hole axial gravitational waves of angular momentum $\ell\ge2$ are described (in units $c=G=2M=1$) by the Regge--Wheeler equation (RWE) \cite{R-W,chand}
\begin{gather}
  [d_x^2 + \w^2 - V(x)]\psi(x,\w)=0\;,\label{RWE}\\
  V(r)=\left(1-\frac{1}{r}\right)
       \left[\frac{\ell(\ell{+}1)}{r^2} - \frac{3}{r^3}\right]\;,\label{V}
\end{gather}
where $x=r+\ln (r{-}1)$ is the tortoise coordinate and $r$ the circumferential radius; $V(r)$ accounts for the Schwarzschild background. The long-range nature of this potential, $V(x)-\nobreak\ell(\ell{+}1)/x^2\sim2\ell(\ell{+}1)\ln x/x^3$ for $x\To\infty$~\cite{centri}, is well known to cause a branch cut in the (retarded) Green's function of (\ref{RWE}) on the negative imaginary axis (NIA) in the $\w$-plane.

For $\c\equiv i\w\downarrow0$, this cut causes an algebraically decaying late-time tail in the gravitational-wave signal \cite{Leaver,Ching-cut}. For moderate $\c$ and $2\le\ell\le4$, it has recently been investigated numerically~\cite{unconv}, leading to a clear conjecture for the large-$\c$ behavior. In a separate development, the strings of quasinormal modes (QNMs) parallel on both sides of the cut and close to it seem to offer clues to the quantum theory, in particular to a calculation of the Bekenstein entropy in loop quantum gravity and to the quantum of area~\cite{hod}. These motivate studying also the branch cut asymptotically, which will turn out to have ample independent interest.

In each $\ell$-sector and in the frequency domain, the RWE is one-dimensional, so the above-mentioned Green's function $\bar{G}(x,y)=\bar{G}(y,x)$ can be written as
\beq
  \bar{G}(x,y;\w)=\frac{f(y,\w)\hp g(x,\w)}{J(\w)}\;,\qquad y<x\;.
\eeql{green}
Here, $f$ solves (\ref{RWE}) with the left outgoing-wave boundary condition (OWC) $f(x{\rightarrow}{-}\infty,\w)\sim1\cdot e^{-i\w x}$ and thus represents waves going into the horizon, while $g(x{\rightarrow}\infty,\w)\sim1\cdot e^{i\w x}$ corresponds to waves going to infinity; $J=gf'-fg'$ is their Wronskian. In the physical region $x\in\mathbb{R}$, these asymptotic definitions are unambiguous only for $\im\w\ge0$, from where the functions are continued analytically. The normalizations of $f,g$ have been fixed for definiteness, but $J$ in the denominator renders (\ref{green}) normalization-independent. Since $V(x{\To}{-}\infty)\sim e^x$, the function $f(x,\w)$ is single-valued in~$\w$. Thus it is intuitively clear, and readily shown~\cite{unconv}, that the branch cut in $\bar{G}$ can be expressed in terms of the one in~$g$.

Focusing on the latter, we define $g_\pm(\w)$ as the continuations from $\re\pm\w>0$, and $\D g\equiv g_+-g_-$. Since $g_\pm(x,-i\c)\sim1\cdot e^{\c x}$ satisfy the \emph{same} (linear, second-order) wave equation, $\D g(x,-i\c)\sim0\cdot e^{\c x}$ is the small solution $\propto\nobreak g(x,+i\c)$. The simple symmetry $g(-\w^*)=g^*(\w)$ renders $\D g$ imaginary, so we introduce the real cut strength $q$ through \cite{Leaver,Alec}
\beq
  \D g(x,-i\c)=iq(\c)\hp g(x,+i\c)\;.
\eeql{def-q}
Since $g$ is defined by the OWC at $x\To\infty$, (\ref{def-q}) defines $q$ not merely $x$-independent, but rather independent of $V(x)$ at any finite $x$: if, say, $V_1(x{>}L) = V_2(x{>}L)$, the corresponding $q_1$ and $q_2$ are identical. Thus, $q$ economically characterizes $\D g$ (and ultimately $\D\bar{G}$), and our task can now be specified as calculating the asymptotics of $q(\c{\To}\infty)$.

\section{WKB solutions}

For $\im\w<0$, in particular on the NIA, the simple asymptotic definition of $g(\w)$ becomes all but meaningless, since it is impossible to distinguish the decaying component $\sim e^{-\c x}$ (to be set to zero) from the `pure' outgoing wave $\sim e^{\c x}$, where this limiting form however has algebraic corrections to all orders in~$V$. One way out is to complement the analytic continuation in frequency with one in \emph{space} \cite{Alec,YTLiu,c-scale}, so that the product $i\w x$ retains a negative real part. However, in terms of $x$, the very equation (\ref{RWE}) is multiple valued, so that the analysis henceforth will proceed in the complex $r$-plane, \emph{viz.},
\beq
  [r^2(r{-}1)^2d_r^2+r(r{-}1)d_r
  -(r{-}1)\{\ell(\ell{+}1)r-3\}+\omega^2r^4]g=0\;.\label{RWE-r}
\eeq

It is possible to impose the OWC for $g_+(r,-i\c)$ [$g_-(r,-i\c)$] directly and stably for $r\To\nobreak-\infty$, and continue the solutions to the physical $r>1$ in the upper (lower) half plane~\cite{Alec}. That is, apart from a trivial overall phase, $g_\pm$ are the \emph{same} solutions as $r$ grows from $-\infty$, until they are prescribed to encircle the singularity $r=0$ in opposite directions. Hence, closer study of this point should shed light on their difference~$\D g$.

At least away from the singularities $r=0,1$ and the anti-Stokes lines (\emph{v.i.}), one expects to have asymptotic expansions $g_\mathrm{a}(r,\w)$ and $g_\mathrm{a}(r,-\w)$, with
\beq
  g_\mathrm{a}(r,\w)\sim{[(r{-}1)e^r]}^{i\w}
          \left\{1+\frac{g_1(r)}{\w}+\frac{g_2(r)}{\w^2}+\cdots\right\},
\eeql{ga}
where the first factor is just a plane wave in the tortoise coordinate. Substitution yields
\begin{align}
  g_1(r)&=\frac{1}{2i}\int_\infty^r\!\frac{ds\,s}{s{-}1}\,V(s)=
  i\biggl[\frac{\ell(\ell{+}1)}{2r}-\frac{3}{4r^2}\biggr]\;,\\
  g_2(r)&=\frac{1}{4}V(r)
    -\frac{1}{8}\biggl[\int_\infty^r\!\frac{ds\,s}{s{-}1}\,V(s)\biggr]^2=
  -\frac{3N}{8r^2}+\frac{\ell(\ell{+}1)-6}{8r^3}+\frac{15}{32r^4}\;,
\end{align}
where $N=8\bigl(\begin{smallmatrix}\ell+2 \\ 4\end{smallmatrix}\bigr)=\frac{4}{3}\nu(\nu{+}1)$ with $\nu\equiv\frac{1}{2}(\ell{-}1)(\ell{+}2)$. That is, these two orders still agree with the upshot of $\w$-expanding the standard WKB expression, though this is no longer true for $g_3$, which involves $V'$ (of course, there are higher-order corrections to WKB as well).

\begin{figure}
\includegraphics[width=5in]{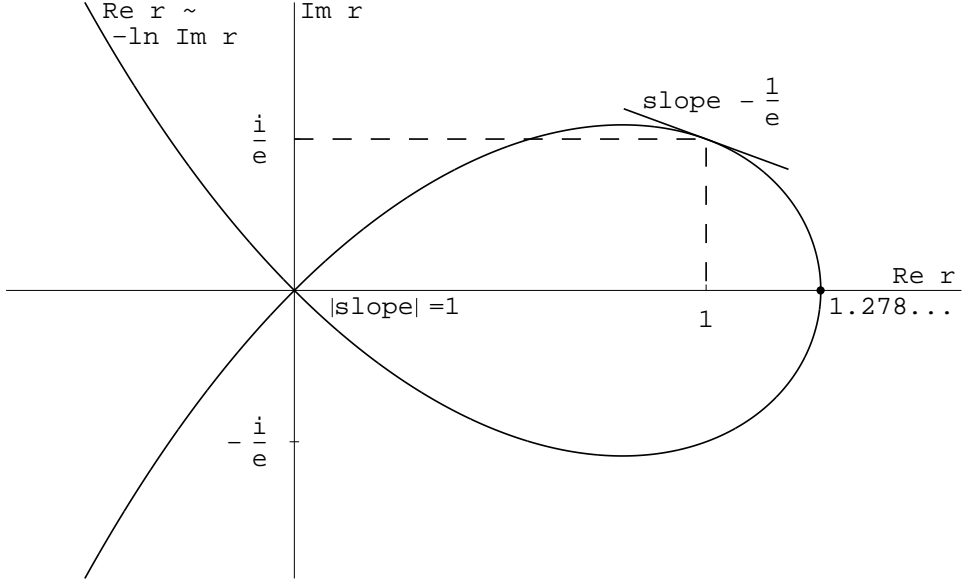}
\caption{Anti-Stokes lines of the RWE for negative imaginary frequencies.}
\label{anti-Stokes}
\end{figure}

The anti-Stokes lines of the RWE on the NIA for this expansion are shown in Fig.~\ref{anti-Stokes}. They are the curves where $|(r{-}1)e^r|=1$ (i.e., $\re x=0$); for $\c\To\infty$, these are the boundaries where the solutions (\ref{ga}) change between exponentially growing and decaying characters.

Thus, the solution which is small for $r\To-\infty$ can be continued to the region including $x>0$, but this only yields the exponentially growing part one knew all along. From there, the solution can\emph{not} be continued back to $x<0$ (where its imaginary part could be identified) because of the Stokes phenomenon. For the latter continuation, we apparently have to pass through the eye of the storm---the black-hole singularity $r=0$~\cite{neitzke}. Near $r=0$, the expansion (\ref{ga}) is not valid, for higher-order terms only become small if $|r|\gg1/\!\sqrt{\c}$. One therefore has to match (possibly different) expansions in terms of $g_\mathrm{a}$ in the regions $\left|\arg r-\pi\right|\le\frac{\pi}{4}$ and $\left|\arg r\right|\le\frac{\pi}{4}$ respectively. This is similar to the connection procedure near classical turning points in Bohr--Sommerfeld quantization. However, near the latter it is merely the asymptotic expansion which breaks down, while the original (Schr\"odinger) wave equation is perfectly regular. In our case, the connection has to be carried out across a singularity of the RWE.

\section{Connection formula}

Series expansion around $r=0$ is standard~\cite{Leaver series}:
\begin{subequations}\label{psi12}
\begin{align}
  \psi_1(r)&=r^3+\frac{6-\ell(\ell{+}1)}{5}\hp r^4+\O(r^5)\;,\label{psi1}\\
  \psi_2(r)&=r^{-1}+\frac{2\nu}{3}+\frac{N}{4}\hp r
             +\frac{\ell(\ell{+}1)N}{12}\hp r^2
             +\frac{\W^2-\w^2}{4}\hp \psi_1(r)\ln r+\O(r^4)\;,\label{psi2}
\end{align}
\end{subequations}%
where both error terms are single-valued, and where we introduced the \emph{algebraically special frequency} $\W=-iN/2$ \cite{Alec,couch}. However, because of the large term $\propto\w^2$ in the RWE (\ref{RWE-r}), higher-order terms in these expansions can only be omitted if $|r^2\w|\ll1$ (cf.\ the given terms of $\psi_2$). Thus, there is no overlap with the region of validity of $g_\mathrm{a}$, where matching could be carried out. Still, the above series for $\psi_{1,2}$ will be useful for comparison, \emph{inter alia} yielding the exact monodromy~\cite{ince} $\psi_2(re^{2\pi i})=\psi_2(r)+i\frac{\pi}{2}(\W^2{-}\w^2)\hp\psi_1(r)$ and $\psi_1(re^{2\pi i})=\psi_1(r)$.

A usable matching solution follows by effectively resumming the large-$\gamma$ parts of higher-order terms in (\ref{psi12}). In practice, it is more convenient to set $r\equiv t/\!\sqrt{\c}$ and sort powers of~$\c$:
\begin{gather}
  \label{R}[t^2d_t^2-td_t-3-t^4]\psi=\frac{1}{\sqrt{\c}}
   [2t^3d_t^2-t^2d_t-\{\ell(\ell{+}1)+3\}t]\psi+\O(\c^{-1})\equiv R(t)\;,\\
  \psi=\psi^{(0)}+\psi^{(1)}+\cdots\;.
\end{gather}
The lowest order follows by equating the rhs of (\ref{R}) to zero:
\begin{subequations}
\begin{align}
  \psi_1^{(0)}(t)&=\frac{4it}{\c^{3/2}}\hp J_1\biggl(\frac{t^2}{2i}\biggr)\;,\\
  \psi_2^{(0)}(t)&=\frac{i\pi\sqrt{\c}}{4}\hp tY_1\biggl(\frac{t^2}{2i}\biggr)\;.
\end{align}
\end{subequations}%
These have not been written in terms of modified Bessel functions, since the subsequent matching is best done on the anti-Stokes lines where $t^2\!/2i$ is real. Subsequently, $\psi_1^{(0)}$ figures as an inhomogeneous term in the equation for $\psi_1^{(1)}$, solved by
\begin{align}
  \psi_1^{(1)}(t)&=\frac{\pi}{4}t\int_0^t\frac{ds}{s^2}
    \left[Y_1\biggl(\frac{t^2}{2i}\biggr)\hp J_1\biggl(\frac{s^2}{2i}\biggr)-
    J_1\biggl(\frac{t^2}{2i}\biggr)\hp
    Y_1\biggl(\frac{s^2}{2i}\biggr)\right]R_1(s)\;,\\
  R_1(s)&=\frac{4i}{\c^2}\left\{2[s^6-\nu s^2]\hp
    J_1\biggl(\frac{s^2}{2i}\biggr)-is^4J_0\biggl(\frac{s^2}{2i}\biggr)\right\}.
    \label{R1}
\end{align}
Note that the occurrence of $Y_1$ does not spoil the analyticity of~$\psi_1^{(1)}$. In particular, for $t\To0$ the above readily reproduces the $t^4$-term found in (\ref{psi1}) by direct expansion. The counterpart for $\psi_2$ reads
\begin{align}
  \psi_2^{(1)}(t)&=\frac{\pi}{4}\hp tY_1\biggl(\frac{t^2}{2i}\biggr)
    \int_0^t\frac{ds}{s^2}J_1\biggl(\frac{s^2}{2i}\biggr)R_2(s)\notag\\
    &\quad+\frac{\pi}{4}\hp tJ_1\biggl(\frac{t^2}{2i}\biggr)\biggl[\int_0^t\!\!ds
    \biggl\{\frac{8i\nu}{\pi s^4}-Y_1\biggl(\frac{s^2}{2i}\biggr)
    \frac{R_2(s)}{s^2}\biggr\}+\frac{8i\nu}{3\pi t^3}\biggr]\;,\label{psi21}\\
  R_2(s)&=\frac{i\pi}{4}\biggl\{2[s^6-\nu s^2]\hp
    Y_1\biggl(\frac{s^2}{2i}\biggr)-is^4Y_0\biggl(\frac{s^2}{2i}\biggr)\biggr\}\;,
\end{align}
where we introduced a `counterterm' in order to keep the second integral finite near $s=0$. Again, one verifies that the leading small-$t$ correction [$\propto t^0$ in (\ref{psi2})] is reproduced correctly. Using the standard branching properties of the $Y_n$ one now finds that, up to this second order, $\psi_2(te^{2\pi i})=\psi_2(t)+i\frac{\pi}{2}\c^2\psi_1(t)$, so that the exact monodromy is approached for $|\w|\gg|\W|$.

In fact, we can follow the transformation of $\psi_{1,2}^{(i)}$ under rotation in more detail by using the standard $J_n(-z)=(-)^nJ_n(z)$ and $Y_n(e^{\pm i\pi}z)=(-)^n[Y_n(z)\pm2iJ_n(z)]$, implying
\begin{subequations}\label{rotate}
\begin{align}
  \psi_1^{(0)}(it)&=-i\psi_1^{(0)}(t)\;,\\
  \psi_1^{(1)}(it)&=\psi_1^{(1)}(t)\;,\\
  \psi_2^{(0)}(it)&=-i\psi_2^{(0)}(t)+\frac{\pi\c^2}{8}\hp\psi_1^{(0)}(t)\;,
    \displaybreak[0]\\
  \psi_2^{(1)}(it)&=\psi_2^{(1)}(t)+i\frac{\pi\c^2}{8}\hp\psi_1^{(1)}(t)\;.
\end{align}
\end{subequations}%
These relations streamline the asymptotic expansion. Namely, look along $\arg t=\frac{\pi}{4}$, where $\psi_{1,2}^{(0)}\hp e^{-3\pi i/4}\in\mathbb{R}$ and $\psi_{1,2}^{(1)}\in\mathbb{R}$, so that their expansions will read
\begin{subequations}\label{match-asy}
\begin{align}
  \psi_1(t)&\sim\frac{4}{\sqrt{\pi\c^3}}\biggl[e^{t^2\!/2}
    \biggl\{1{+}\frac{\alpha}{\sqrt{\c}}\biggr\}+e^{-t^2\!/2}
    \biggl\{-i{+}\frac{\alpha^*}{\sqrt{\c}}\biggr\}+\O(t^{-2})
           +\O\biggl(\frac{t^3}{\sqrt{\c}}\biggr)+\O(\c^{-1})\biggr]\,,
           \displaybreak[0]\\
  \psi_2(t)&\sim\frac{\sqrt{\pi\c}}{4}\biggl[e^{t^2\!/2}
    \biggl\{-i{+}\frac{\beta}{\sqrt{\c}}\biggr\}+e^{-t^2\!/2}
    \biggl\{1{+}\frac{\beta^*}{\sqrt{\c}}\biggr\}+\O(t^{-2})
           +\O\biggl(\frac{t^3}{\sqrt{\c}}\biggr)+\O(\c^{-1})\biggr]\,.
\end{align}
\end{subequations}%
The occurrence of $\O(t^3\!/\!\sqrt{\c})$ [which however does \emph{not} generate additional $t^0\!/\!\sqrt{\c}$-terms] means that these hold for $1\ll t\ll\c^{1/6}$ or $\c^{-1/2}\ll r\ll\c^{-1/3}$, conveniently handled as a double asymptotic expansion in $t$ and $\lambda\equiv\c/t^6$. The power-law asymptotic corrections in $t$ follow directly from the RWE; the real nonlocal information is contained in~$\alpha,\beta$.

We can use the general rules (\ref{rotate}) to transform (\ref{match-asy}) to $\arg t=\frac{3\pi}{4}$, and demand consistency for the ($\propto e^{-t^2\!/2}$) part that dominates for $\frac{\pi}{4}<\arg t<\frac{3\pi}{4}$. This leads to $\alpha\in\mathbb{R}$ and $\im\beta=-\alpha$. By subsequently demanding consistency also for the combination of $\psi_1$ and $\psi_2$ which is minimal in the same sector, one finds
\beq
  \beta=-(2{+}i)\alpha\;.
\eeq
Thus, this algebraic exercise circumvented directly expanding the integrals (\ref{psi21}) for~$\psi_2^{(1)}$.

\section{Matching}

For $\left|\arg t-\pi\right|<\frac{3\pi}{4}$ one can do the analogous expansion of $g_+(t,-i\c)=g_\mathrm{a}(t,-i\c)$ in~(\ref{ga}). Again, one finds a Gaussian form if $t\ll\c^{1/6}$~\cite{precond}:
\beq\begin{split}
  g_+(t,-i\c)&\sim e^{i\pi\c}e^{-t^2\!/2}\biggl[1+\frac{3}{4t^2}-\frac{15}{32t^4}
  -\frac{t^3}{3\sqrt{\c}}-\frac{t}{4\sqrt{\c}}+\frac{5{-}16\ell(\ell{+}1)}{32t\sqrt{\c}}
  +\frac{t^6}{18\c}+\mathrm{h.o.t.}\biggr]\;;
\end{split}\eeql{ga-asy}
the first factor comes from encircling the horizon $r=1$. The leading corrections verify the consistency of expansions (\ref{match-asy}) and (\ref{ga-asy}), obtained in very different ways; all that matters for the matching is $[\cdots]=1+0\cdot t^0\!/\!\sqrt{\c}+\cdots$.

Comparison shows that
\beq
  g_+e^{-i\pi\c}=
  \frac{\sqrt{\pi}}{8}\bigl(3i\c^{3/2}+(2{-}3i)\alpha\c+\O(\sqrt{\c})\bigr)\psi_1
  +\frac{2}{\sqrt{\pi}}\biggl(-\frac{1}{\sqrt{\c}}+\frac{\alpha}{\c}
                              +\O(\c^{-3/2})\biggr)\psi_2\;,
\eeq
which can be matched back to solutions in terms of $g_\mathrm{a}$ on $\arg t=\frac{\pi}{4}$, yielding
\beq
  g_+(-i\c)=g_\mathrm{a}(-i\c)
    +2e^{2\pi i\c}\biggl(i+\frac{\alpha}{\sqrt{\c}}\biggr)g_\mathrm{a}(+i\c)
\eeql{g-match}
in the region (bounded by anti-Stokes lines) including $r=1$. Hence, in particular in the physical part $x<0$ of that region, one has
\beq\begin{split}
  \D g(-i\c)&=2i\im g_+(-i\c)=4i\biggl[\cos(2\pi\c)
    +\frac{\alpha}{\sqrt{\c}}\sin(2\pi\c)\biggr]g(+i\c)\\
  &\equiv iq(\c)g(+i\c)\;.
\end{split}\eeql{delta-g}

Already, the numerically observed asymptotics~\cite{unconv} of $q(\c)$ have been confirmed. For the corrections, it remains to calculate $\alpha$ in closed form. Let us start with the $\nu$-dependent term in (\ref{R1}), which clearly cannot be combined with the other two. Straightforward manipulations yield its contribution to $\psi_1^{(1)}$ as ($\arg t=\frac{\pi}{4}$)
\beq
  \psi_1^{(1)}(t)\sim\frac{\sqrt{8\pi}}{\c^2}\nu\biggl[
    \cos\Bigl(\frac{t^2}{2i}{+}\frac{\pi}{4}\Bigr)
    \int_0^\infty\!\frac{dz}{\sqrt{z}}\,J_1(z)\hp Y_1(z)
    +\cos\Bigl(\frac{t^2}{2i}{+}\frac{3\pi}{4}\Bigr)
    \int_0^\infty\!\frac{dz}{\sqrt{z}}\,J_1^2(z)\biggr]\;.
\eeq
By considering the asymptotics of $\int_0^K(dz/\!\sqrt{z})\hp J_1(z)\hp H_1^{(1)}(z)$ ($H$ are Hankel functions) in the upper-half $K$-plane [again using the formula for $Y_1(e^{i\pi}z)$ above (\ref{rotate})], one convinces oneself that in fact $\int_0^\infty(dz/\!\sqrt{z})\hp J_1(z)\hp Y_1(z)=-\int_0^\infty(dz/\!\sqrt{z})\hp J_1^2(z)$, which is also necessary for this contribution to $\alpha$ to be real. The latter integral is tabulated as $\int_0^\infty(dz/\!\sqrt{z})\hp J_1^2(z)=\Gamma\bigl(\frac{1}{4}\bigr)^4\!/12\pi^{5/2}$~\cite{table}. Thus, the present contribution to $\alpha$ reads
\beq
  \alpha_1=-\frac{\nu\Gamma\bigl(\frac{1}{4}\bigr)^4}{24\pi^{3/2}}\;.
\eeql{a1}
The last term of (\ref{R1}) analogously leads to the integrals $\int_0^\infty\!dz\,\sqrt{z}\hp J_1(z)\hp J_0(z)=-\int_0^\infty\!dz\,\sqrt{z}\hp Y_1(z)\hp J_0(z)=\Gamma\bigl(\frac{1}{4}\bigr)^4\!/16\pi^{5/2}$, for a contribution
\beq
  \alpha_2=\frac{\Gamma\bigl(\frac{1}{4}\bigr)^4}{32\pi^{3/2}}\;.
\eeql{a2}
However, the first term of (\ref{R1}), with its higher power of $s$, leads to diverging integrals:
\begin{align}
  \int_0^K\!\!dz\,z^{3/2}J_1^2(z)&\sim\frac{2}{3\pi}K^{3/2}
    +\frac{1}{2\pi}\sqrt{K}\cos(2K)+c_3+\O(K^{-1/2})\;,\\
  \int_0^K\!\!dz\,z^{3/2}J_1(z)\hp Y_1(z)&\sim\frac{1}{2\pi}\sqrt{K}\sin(2K)+d_3
    +\O(K^{-1/2})\;.
\end{align}
\emph{If} $d_3=-c_3$, these lead to a real contribution
\beq
  \alpha_3=-2\pi c_3\;.
\eeql{a3}
Unfortunately, the general $\int_0^K\!dz\,z^{3/2}J_1^2(z)= (K^{9/2}\!/18)\hp _2F_3(\frac{3}{2},\frac{9}{4};2,3,\frac{13}{4};-K^2)$ does not help directly, since not enough seems to be known about the asymptotics of~$_2F_3$. Instead, one can proceed as follows: $d_3=-c_3$ can again be proven by studying $\int_0^K\!dz\,z^{3/2}J_1(z)\hp H_1^{(1)}(z)$ in the upper-half plane. It is then logical to also consider $\int_0^K\!dz\,z^{3/2}H_1^{(1)}(z)^2$, in which one can take $K\To i\infty$. One finds
\beq
  c_3=\frac{\sqrt{2}}{\pi^2}\int_0^\infty\!\!dw\,w^{3/2}K_1^2(w)
     =\frac{5\Gamma\bigl(\frac{1}{4}\bigr)^4}{192\pi^{5/2}}\;.
\eeql{c3}
The rest is straightforward: (\ref{a1}), (\ref{a2}), and (\ref{a3}) with (\ref{c3}) can be added and substituted into (\ref{delta-g}), from which one can read off our final answer
\beq
  q(\c)\sim4\cos(2\pi\c)+\frac{\Gamma\bigl(\frac{1}{4}\bigr)^4}{12\pi^{3/2}}
  \hp[1-\ell(\ell{+}1)]\hp\frac{\sin(2\pi\c)}{\sqrt{\c}}+\O(\c^{-1})\;,
\eeql{q-res}
where $\Gamma\bigl(\frac{1}{4}\bigr)^4\!/12\pi^{3/2}=2.586\ldots$ The \emph{result} for the leading term has appeared before in Ref.~\cite{unconv}; preliminary results from a transmission-amplitude calculation seem to support the form for the corrections, in particular the $\ell$-dependence~\cite{AN}.

\section{Discussion}

The above analysis, for $r,\c\To\infty$, is the third instance where the mathematics of the RWE near $r=0$ has been seen to influence the goings-on in our universe $r>1$. The first instance is the question of `anomalous' vs `miraculous' waves outgoing into the horizon, i.e., $r\downarrow1$ and $2\c\in\mathbb{N}$ (\cite{Alec}, Section~VI). The second instance is the dynamics at and around the algebraically special frequency $\c=N/2$, where the RWE has closed-form solutions, whose global behavior in the $r$-plane can therefore be traced~\cite{Alec,unconv}.

Besides the axial waves described by the RWE (\ref{RWE})--(\ref{V}) there are also polar waves, described by the Zerilli equation~\cite{Zerilli}. However, the latter's `intertwining' or supersymmetry relation to the RWE yields its cut strength as $\tilde{q}(\c)=[(N/2{-}\c)/(N/2{+}\c)]q(\c)$~\cite{Alec}. Thus, $q$ and $\tilde{q}$ agree up to an overall sign, plus $\O(\c^{-1})$ corrections immaterial to~(\ref{q-res}).

For an outlook, the first obvious item is the numerical verification of (\ref{q-res}), in particular of the sub-leading correction. A brute-force increase of numerical precision in the existing method is unlikely to suffice, especially for $\ell\ge3$. More promising is to calculate $g_\pm(-i\c)$ directly on the NIA, instead of through extrapolation to this axis. In the series for $g(\w)$ \cite{unconv,Leaver series}, the problem for $\re\w\To0$ is not convergence but rather obtaining the individual (irregular-hypergeometric) terms reliably. Working this out should be mainly a matter of time, but it remains to be seen if it would sufficiently extend the range of validity in~$\c$.

Related to this, it is worth re-emphasizing~\cite{Alec} that the present method involves two conceptually separate steps: (a)~the continuation in~$r$, stabilizing the OWC at infinity in the lower-half $\w$-plane, leading to a well-defined computational problem, and (b)~the asymptotics, by which one can actually solve this problem analytically for large~$\c$. The second step is optional~\cite{YTLiu}, and numerical integration of $g_+$ from $r=-\infty$ (combined with standard series solutions for~$f$) should soon open up the third $\w$-quadrant (i.e., behind the cut) to direct exploration, especially for at most moderate damping.

It would also be interesting if this work could be compared to the closed-form expression for $q(\c)$ in (31)--(33) of Ref.~\cite{Leaver}, through the coefficients $d_L^{(\nu)}$ of an expansion $g\propto\sum_{L=-\infty}^\infty d_L^{(\nu)}u_{L+\nu}$; here, $u_{L+\nu}$ are Coulomb wave functions. As usual, the $d_L^{(\nu)}$ satisfy a three-term recursion relation (given in a simplified form, which can be made purely real on the NIA, in Section~VI~F of~\cite{Leaver series}); $0\le\nu<1$ is to be chosen such that $d_L^{(\nu)}$ is the minimal solution to this relation for both $L\To-\infty$ and~$\infty$. The adiabatic Ansatz of slowly varying $d_{L+1}^{(\nu)}/d_L^{(\nu)}$ readily yields asymptotic solutions for $\c\To\infty$, except near $L=\pm\c$ and $L=0$. Following Ref.~\cite{motl}, one can try to develop connection formulas near these three points, which would determine $\nu$ analytically. At the `turning points' $L=\pm\c$, the three-term recursion asymptotically degenerates into a \emph{two}-term one, and the connection proceeds exactly as in~\cite{motl}. Near $L=0$, however, all three terms are of the same order (in~$\c$), and the recursion remains in the form of analytically intractable continued fractions (i.e., without simplifying to products). Thus, this route for now seems unfeasible.

At least to leading order, the method of continuation through the vicinity of $r=0$ has meanwhile independently been used to calculate the high-damping QNM frequencies~\cite{neitzke}. On the one hand, this yields more information on the QNM \emph{wave functions} than the continued-fraction technique~\cite{motl}. On the other, this clearly establishes a relation to the present problem of the branch-cut strength~$q(\c)$. Thus one can expect also \emph{numerical} results for these two high-damping aspects of the RWE to bear on each other, and an extension to other black-hole models. Conversely, this work's progress on the corrections should help finding the $\O(|\w|^{-1/2})$ terms for the asymptotic QNM frequencies. Meanwhile, this has indeed been possible; see the Appendix.

In particular, these developments render it urgent to study the branch cut of the RWE's Green's function also for other values of the spin~$s$, where (\ref{V}) corresponds to $s=2$. Notably, the highly damped electromagnetic ($s=1$) QNMs are predicted~\cite{neitzke,motl} to approach the NIA asymptotically so that one expects an even closer, though yet unknown, relation to the branch cut in that case.


\begin{acknowledgments}
The work of the main text was performed in summer 2001 during a sabbatical as a C.N. Yang Visiting Fellow at the Chinese University of Hong Kong. The recent activity in this field has prompted me to publish my original notes. I thank V.~Cardoso, A.~Neitzke, and K.~Young for fruitful discussions.
\end{acknowledgments}

\appendix

\section{High-damping QNMs}

For frequencies close to the NIA, it should certainly be possible to analytically continue~(\ref{g-match}) and find those $\c$ for which $g_+$ also satisfies the OWC into the horizon, i.e.\ $g_+(\w)\propto\nobreak f(\w)$. These should identify the highly damped QNMs. Here, $f$ is characterized by its monodromy around the horizon:
\beq
  f\bm{(}(r{-}1)e^{2\pi i},\w\bm{)}=e^{2\pi\w}f(r{-}1,\w)\;.
\eeql{mono}
Inspecting~(\ref{ga}), it seems that `$f(\w)=g_\mathrm{a}(-\w)$', but this is deceptive. While it \emph{is} true that $f(-\w)=g_\mathrm{a}(\w)$ in the lower-half $\w$-plane, the asymptotic nature of the large (near $r=1$) solution $g_\mathrm{a}(-\w)$ means that naive rotation of $r-1$ only confirms the desired monodromy to dominant order, whereas $f$ is required to obey it exactly.

The solution is to, in the spirit of \cite{neitzke,AN}, do the rotation along the anti-Stokes contour in Fig.~\ref{anti-Stokes}, where neither solution dominates the other (skipping $r=0$ on the inside). By the normalization of $f$ one knows the dominant component, so we take the Ansatz
\beq
  f(-i\c)=g_\mathrm{a}(i\c)+c(\c)\hp g_\mathrm{a}(-i\c)\;.
\eeql{f-c}
Continuing this from the physical $x=0$ to $1/\!\sqrt{\c}\ll|r|\ll1$ with $\arg r=\frac{\pi}{4}$, one finds
\beq\begin{split}
  f(-i\c)&\sim e^{-i\pi\c}e^{t^2\!/2}+c(\c)\hp e^{i\pi\c}e^{-t^2\!/2}\\
  &\sim\frac{\sqrt{\pi\c^3}}{8}\left[e^{-i\pi\c}
    \left\{1+\frac{(i{-}2)\a}{\sqrt{\c}}\right\}
    +c(\c)\hp e^{i\pi\c}\left\{i+\frac{(2{+}i)\a}{\sqrt{\c}}\right\}\right]\psi_1\\
  &\quad+\frac{2}{\sqrt{\pi\c}}\left[e^{-i\pi\c}\left\{i-\frac{\a}{\sqrt{\c}}\right\}
    +c(\c)\hp e^{i\pi\c}\left\{1+\frac{\a}{\sqrt{\c}}\right\}\right]\psi_2\;;
\end{split}\eeq
the second line followed by comparison with (\ref{match-asy}). As before, (\ref{rotate}) makes quick work of continuing this to $\arg r=-\frac{\pi}{4}$, where it can be matched back to a combination of $g_\mathrm{a}(\pm i\c)$. One thus finds $f\bm{(}(r{-}1)e^{2\pi i},-i\c\bm{)}=e^{-2\pi i\c}g_\mathrm{a}(r{-}1,i\c)+[2i(1{-}\a/\!\sqrt{\c})+c(\c)\hp e^{2\pi i\c}]\*g_\mathrm{a}(r{-}1,-i\c)$. Indeed $f$ as in (\ref{f-c}) obeys (\ref{mono}) to dominant order for any $c$, while the subdominant ($\propto c$) term by itself has the opposite monodromy, corresponding to incoming waves---both as stipulated above. Equation~(\ref{mono}) holds exactly for
\beq
  c(\c)\sim-\frac{1-\a/\!\sqrt{\c}}{\sin(2\pi\c)}\;.
\eeql{c-res}
As required, $c(\c{\in}\mathbb{R})\in\mathbb{R}$, since the exponential tail of $V(x{\To}{-}\infty)$ does not generate a branch cut in~$f(\w)$. Also the poles for $2\c\in\mathbb{N}$ are not surprising, since the RWE is known to have such \emph{anomalous points} for $2\c=1,2,\ldots$, with exactly \emph{one} exception at $2\c=N$~\cite{Alec}; clearly, the latter is beyond the reach of the present asymptotics. Expressing $f(\pm\w)$, $g(\pm\w)$ not through $g_\mathrm{a}$ but in terms of each other, the $S$-matrix could be read off; cf.~\cite{AN} for the leading order.

Combining (\ref{g-match}), (\ref{f-c}), and (\ref{c-res}), all that remains is to asymptotically solve
\beq\begin{split}
  0=J(-i\c)&\propto\sin(2\pi\c)+2e^{2\pi i\c}\biggl(i+\frac{\a}{\sqrt{\c}}\biggr)
    \biggl(1-\frac{\a}{\sqrt{\c}}\biggr)\\
   &\approx\sin(2\pi\c)+2ie^{2\pi i\c}\biggl(1-\frac{(1{+}i)\a}{\sqrt{\c}}\biggr)\;.
\end{split}\eeq
Reexpressing the answer in terms of $\w$ and substituting $\a$ from the main text, one obtains
\beq
  \w_n=\frac{\ln3}{4\pi}-\left({\ts\frac{n}{2}{+}\frac{1}{4}}\right)i+
  \frac{\sqrt{2}\Gamma\bigl(\frac{1}{4}\bigr)^4}{144\pi^{5/2}}\hp(1{+}i)\hp
  \frac{\ell(\ell{+}1)-1}{\sqrt{n}}+\O(n^{-1})\;,
\eeql{om-n}
where the prefactor of the correction evaluates to $0.097007\ldots$ The $\w_n$ have automatically come out on the physical sheet of $g_+$; contrast Ref.~\cite{motl}, which is not sheet-specific. Agreement with (29)--(30) in Ref.~\cite{nollert} is excellent; as anticipated in the Discussion, this numerical confirmation of the value of $\a$ via the \emph{QNMs} greatly supports (\ref{q-res}) for the \emph{cut}. To my knowledge, this is the first time that the correction to the highly damped Schwarzschild QNM frequencies has been calculated analytically.



\begin{thebibliography}{99}
\bibitem{centri}Since the centrifugal potential $\propto x^{-2}$ does not scatter, it does not contribute to the cut and hence has been subtracted.
\end{thebibliography}
\end{document}